\documentclass[oldversion]{aa}
\usepackage{graphicx}
\begin{document}
%
   \title{A comprehensive examination of the $\epsilon$ Eri system
\thanks{Based on observations collected at the European Southern Observatory, 
Chile (ESO No.\ 080.C-0598).}
}

   \subtitle{Verification of a 4 micron narrow-band high-contrast imaging approach for planet searches}

   \author{Markus Janson\inst{1} \and
          Sabine Reffert\inst{2} \and
          Wolfgang Brandner\inst{1} \and
	  Thomas Henning\inst{1} \and
	  Rainer Lenzen\inst{1} \and
	  Stefan Hippler\inst{1}
          }

   \offprints{Markus Janson}

   \institute{Max-Planck-Institut f\"ur Astronomie, K\"onigstuhl 17, D-69117 Heidelberg, Germany\\
              \email{janson@mpia.de, brandner@mpia.de, henning@mpia.de, lenzen@mpia.de, hippler@mpia.de}
         \and
             Landessternwarte, K\"onigstuhl 12, D-69117 Heidelberg, Germany\\
	      \email{sreffert@lsw.uni-heidelberg.de}
             }

   \date{Received ---; accepted ---}

   \abstract{
Due to its proximity, youth, and solar-like characteristics with a spectral type of K2V, $\epsilon$ Eri is one of the most extensively studied systems in an extrasolar planet context. Based on radial velocity, astrometry, and studies of the structure of its circumstellar debris disk, at least two planetary companion candidates to $\epsilon$ Eri have been inferred in the literature ($\epsilon$ Eri b, $\epsilon$ Eri c). Some of these methods also hint at additional companions residing in the system. Here we present a new adaptive optics assisted high-contrast imaging approach that takes advantage of the favourable planet spectral energy distribution at 4 $\mu$m, using narrow-band angular differential imaging to provide an improved contrast at small and intermediate separations from the star. We use this method to search for planets at orbits intermediate between $\epsilon$ Eri b (3.4 AU) and $\epsilon$ Eri c (40 AU). The method is described in detail, and important issues related to the detectability of planets such as the age of $\epsilon$ Eri and constraints from indirect measurements are discussed. The non-detection of companion candidates provides stringent upper limits for the masses of additional planets. Using a combination of the existing dynamic and imaging data, we exclude the presence of any planetary companion more massive than 3 $M_{\rm jup}$ anywhere in the $\epsilon$ Eri system. Specifically, with regards to the possible residual linear radial velocity trend, we find that it is unlikely to correspond to a real physical companion if the system is as young as 200 Myr, whereas if it is as old as 800 Myr, there is an allowed semi-major axis range between about 8.5 and 25 AU.
   
}{}{}{}{}

   \maketitle
%

\section{Introduction}
\label{sec_intro}

With its distance of 3.2 pc and spectral type K2V, $\epsilon$ Eri is the most nearby solar-like single star. Largely due to this fact, it has been targeted by several attempts to indirectly or directly detect planetary companions. So far, two exoplanet candidates have been reported as companions to $\epsilon$ Eri -- $\epsilon$ Eri b (1.55 $M_{\rm jup}$, $a=3.4$ AU), which was inferred from a periodic radial velocity signal by Hatzes et al. (2000) and subsequently also from astrometric motion of the primary (Benedict et al. 2006), and $\epsilon$ Eri c (0.1 $M_{\rm jup}$, 40 AU), which results from an interpretation of observed irregularities in the dust disk surrounding the primary (Quillen \& Thorndike, 2002). Targeted attempts have been made to directly image both $\epsilon$ Eri b (Janson et al. 2007) and $\epsilon$ Eri c (Macintosh et al. 2003, Marengo et al. 2006), but did not yield any detections. $\epsilon$ Eri has also been observed as a part of large-scale near-infrared adaptive optics surveys, such as Lafreniere et al. (2007), and Biller et al. (2007).

Typically, direct imaging surveys from the ground are performed in narrow-band filters around 1.6 $\mu$m, since atmospheric opacities cause cool objects (giant planets and brown dwarfs) with temperatures below 800 K to emit a large fraction of their H-band flux just shortward of the CH$_4$ absorption feature at 1.6 $\mu$m, and since this feature also allows for spectral differential imaging (SDI) to enhance the image contrast. However, the primary-to-secondary brightness contrast is generally smaller at longer wavelengths, such that an improved performance can be gained for some types of high-contrast systems. This was demonstrated with an imaging survey in the L'-band of 22 young, nearby stars by Kasper et al. (2007). The brightness difference is even smaller in M-band (see e.g. Hinz et al. 2006), but here the noise contribution of the thermal background is also significantly higher, making it observationally challenging. Here we examine an intermediate approach, where we use a narrow-band filter towards the red end of the L'-band range, which simultaneously allows for both good brightness contrast and observational simplicity. We test this technique on the $\epsilon$ Eri system to search for planetary companions at separations intermediate between $\epsilon$ Eri b and $\epsilon$ Eri c.

The paper is organized as follows: In Sect. \ref{sec_obsstrat}, we describe the basic principles of the observing strategy chosen, and in Sect. \ref{sec_obs} we describe the actual observations performed, in terms of the techniques used and the observing conditions and corresponding instrument performance. This is followed in Sect. \ref{sec_datared} by a description of the data reduction, and the results as given in Sect. \ref{sec_contrast}. Our choice of filter is further motivated in Sect. \ref{sec_comparison}, where it is also explained over which parameter range the method should be preferable to other common methods in high-contrast imaging. In  Sect. \ref{sec_age}, we discuss the controversial issue of the age of $\epsilon$ Eri, which has fundamental implications for the detectability of planetary companions in the system. In Sect. \ref{sec_masslimits} the brightness contrast is translated into a mass contrast. We examine the possible clues for an intermediate companion in Sect. \ref{sec_dyn}. Finally, we summarize what these issues mean in terms of mass-equivalent detection limits for exoplanet companions to $\epsilon$ Eri in Sect. \ref{sec_limits}, and we conclude in Sect. \ref{sec_conclusion}.

\section{Observing strategy}
\label{sec_obsstrat}

Both the optimal choice of strategy, as well as the actual detection limits for a given exoplanet survey, depend crucially on the spectral energy distribution of planets in the mass- and age ranges of the sample. Since we lack information about the actual spectral distributions of young planets from observations, we need to rely on spectral models. For this purpose we use the Burrows et al. (2003) models for relative spectral distributions, and the Baraffe et al. (2003) models for the absolute brightness within a given filter. Similar strategies were applied in, e.g., Janson et al. (2007) and Apai et al. (2008). Here we develop these methods somewhat further. 

The idea of choosing an optimal observing strategy from spectral models is based on the fact that the spectral energy distributions of planets and low-mass brown dwarfs are expected to be extremely non-flat (more precisely, deviate strongly from a blackbody spectrum), with significant flux windows side by side with deep absorption features in the near-infrared range. Optimization is achieved by choosing a spectral range that simultaneously contains a maximal amount of flux from the planet and a minimal amount of flux from the star -- and, in addition, a minimal amount of flux from the thermal background. Since different types of companions will have different spectral properties, the optimization needs to be done with consideration to what is actually being sought after. As a general principle, a narrow-band filter that contains a relatively large fraction of the planetary flux within the wide band that encompasses its spectral range will be preferable to the broad-band itself for the observations. Thus, for instance, the $F1$ narrow-band of SDI is vastly preferable to H-band for this purpose. This issue warrants emphasis, because intuition might suggest the opposite -- for instance, in the case of an isolated faint object, the $SNR$ will be limited by the amount of photons received from the object itself, which can be maximized by maximizing the spectral range of the observations. In the case of PSF-limited high-contrast imaging, however, the interesting signal is provided by the secondary, but the noise is provided by the primary. Hence increasing the spectral range from, e.g., $F1$ to $H$, will add little to the signal, but plenty to the noise, leading to a decrease of the effective $SNR$.

Given a spectral distribution for a certain object, we can determine a magnitude correction $\Delta _{\rm mag}$ between a narrow-band filter ($F_{\rm NB}$) and its encompassing broad-band filter ($F_{\rm BB}$) by simply calculating:

\begin{equation}
\Delta _{\rm mag} = -2.5*{\rm log}_{10}(\bar{f}_{\rm NB} / \bar{f}_{\rm BB})
\end{equation}

where $\bar{f}_{\rm NB}$ is the average flux density in the narrow-band filter, and $\bar{f}_{\rm BB}$ is the same in the broad-band filter. The flux densities are acquired from Burrows et al. (2003). Under the assumption that the stellar spectral features are negligible, $\Delta _{\rm mag}$ can be added to the absolute magnitude of the hypothetical planet under study as given by the Baraffe et al. (2003) models, and compared to the known absolute magnitude of the primary, in order to estimate the physical contrast corresponding to the relevant filter. The contrast must then be compared to the instrumental contrast that can be achieved corresponding to that particular filter. The instrumental contrast depends on a number of factors -- for instance, due to the fact that Strehl ratio increases monotonically with wavelength, the speckle-limited instrumental contrast will improve with increasing wavelength. On the other hand, the thermal background also increases with wavelength, and so the background-limited instrumental contrast will decrease with increasing wavelength. Likewise, the diffraction limit of the telescope increases with increasing wavelength, which can decrease the instrumental contrast at small separation. An additional difference arises from the fact that different techniques are applicable in different filters. For instance, SDI, which improves most PSF-related instrumental contrast is well applicable in H-band, but not in L'-band (at least with presently existing instrumentation). These effects are most efficiently taken into account by comparing actual measured instrumental contrasts, acquired under fair comparative circumstances. Prior to observing $\epsilon$ Eri, we determined that NB4.05 would be the best filter based on the contrast curves achieved in L'-band (see Kasper et al. 2007), but in Sect. \ref{sec_comparison} we will use the more illustrative (and accurate) approach of deriving detection limits using the actual measured contrast curve as presented in Sect. \ref{sec_contrast}.

\section{Observations}
\label{sec_obs}

The target $\epsilon$ Eri was observed at two neighbouring epochs, the first on Dec 17 and the second on Dec 28, 2007. These adaptive optics (AO) observations were performed in service mode with the NACO instrument at the VLT on Paranal, Chile. For the purpose of optimizing the star-planet contrast, the NB4.05 filter (a narrow-band filter centered at 4.05 $\mu$m with a bandwidth of 0.02 $\mu$m) was used for the imaging, this choice is motivated in Sect. \ref{sec_comparison}. The pixel scale was 27 mas/pixel, which corresponds to a field of view of about 27$''$ by 27$''$. A neutral density filter which reduces the transmission by a factor 50 was used during the acquisition in order to avoid saturation in the acquisition image. This filter was removed during the actual observations, which led to saturation of the detector by about a factor 10 at the center of the stellar PSF, and the saturation extended out to 0.2$''$, as had been estimated beforehand. This does not negatively affect the result, aside from the fact that a few minutes need to be spent on empty read-outs at the end of the run in order to ensure that no memory effects will remain on the detector for subsequent observing runs. Angular differential imaging (ADI), as applied in e.g. Janson et al. (2007), was used for this run as well, since it allows for efficient PSF subtraction and yields easily identifiable companion signatures. This was implemented by performing the first half of the observations on each occasion with an instrument rotator orientation of $0^{\circ}$ (with respect to North in the sky), and the other half with an orientation of $33^{\circ}$. A jittering scheme was performed to allow for efficient subtraction of the strong background that can be expected around 4 $\mu$m. Integration times and other observation parameters are given in Table \ref{nbtab1}, where these values are also compared to the conditions of epoch 4 in the multi-epoch observing campaign of $\epsilon$ Eri using NACO-SDI (Janson et al. 2007). 

\begin{table}[htb]
   \centering
\caption[]{Observational conditions for the two runs, compared with the conditions of epoch 4 for the NACO-SDI $\epsilon$ Eri campaign for reference (Janson et al. 2007).}
         \label{nbtab1}
\begin{tabular}{llll}
            \noalign{\smallskip}
            \hline
            \hline
            \noalign{\smallskip}
                 ~ & NB4.05, 1   & NB4.05, 2   & SDI \\
            \noalign{\smallskip}
            \hline
            \noalign{\smallskip}
Main date          & 17 Dec 2007 & 28 Dec 2007 & 1 Jan 2006 \\
Mean seeing$^a$    & 1.42$''$    & 0.65$''$    & 0.82$''$   \\
Mean Strehl$^b$    & 84.3\%      & 87.4\%      & 33.8\%     \\
Mean humidity      & 6\%         & 32\%        & 38\%       \\
Coherence time$^a$ & 2.6 ms      & 3.4 ms      & 3.8 ms     \\
Frames             & 19          & 19          & 52         \\
(per angle)        &      ~      &     ~       &       ~    \\
DIT                & 1.0 s       & 1.0 s       & 1.0 s      \\
NDIT               & 61          & 61          & 86         \\
Tot. time          & 1159 s      & 1159 s      & 4472 s     \\
(per angle)        &      ~      &     ~       &       ~    \\

            \noalign{\smallskip}
            \hline
\end{tabular}
\begin{list}{}{}
\item[$^{\mathrm{a}}$] Values given by the atmospheric seeing monitor at a wavelength of 500 nm.
\item[$^{\mathrm{b}}$] Strehl ratio given by the AO system, rescaled to the observing wavelengths.
\end{list}
\end{table}

Although the atmospheric conditions given by the on-site atmospheric monitor were quite different for the two nights (see Table \ref{nbtab1}), the average quality of the wavefront correction, quantified by the Strehl ratio as given by the AO system itself, was remarkably similar. The exquisite Strehl ratio of well above 80 \% in all cases results from the relatively long wavelengths used. For reference, observations performed with NACO-SDI at 1.6 $\mu$m using the same AO system rarely yield a Strehl ratio above 30 \%. The measured FWHM of the PSF in the non-saturated acquisition image, as given by the 'imexamine' routine in IRAF, was 119 mas.

\section{Data reduction}
\label{sec_datared}

The data reduction was performed in a similar manner as discussed in Kasper et al. (2007), for the part of the sample presented there for which ADI was used. The basic steps of flat fielding, bad pixel removal and background subtraction were performed by the ESO 'jitter' routine in the automatic pipeline with excellent results. The remaining steps of ADI, combination of the images from the two different nights, and analysis of the data quality, were performed with a dedicated IDL pipeline specifically written for this purpose. For the ADI, the $0^{\circ}$ and $33^{\circ}$ images were shifted to a common position using bilinear interpolation, where the relative shift was determined through cross-correlation over the whole image range. The images were also shifted so that the center of the stellar PSF corresponds to the center of the frame, where the absolute shift was determined by calculating the flux-weighted centroid in the image and taking this as the saturated stellar PSF center (since $\epsilon$ Eri is the only bright object in the field). Unsharp-masking was applied in order to remove features of low spatial frequency (stellar halo, background) without affecting higher-frequency features such as the PSF cores of actual companions. For this purpose, image copies convolved with a Gaussian kernel of FWHM 500 mas were produced and subtracted from the original images. Subsequently, the actual subtraction of $0^{\circ}$ and $33^{\circ}$ images was performed, followed by co-addition of the data from the two different nights. Since the uncertainty in the pixel scale is on the order of 0.1 mas/pixel (see e.g. Neuh\"auser et al. 2005), at the largest separations probed there can be deviations of about 50-60 mas, i.e. about half the FWHM. Hence, any object detected in this range could be expected to appear somewhat extended.

Uncertainty levels were estimated in the same way as in Janson et al. (2008), by calculating the pixel-value standard deviation in one-pixel annuli at every separation from the star. For translating the uncertainties into detectable star-planet brightness contrasts, it is necessary to estimate the brightness of the primary star, which can not be done on the basis of the science images themselves, as they are saturated. Instead, this was performed by estimating the brightness (in terms of the PSF peak value) in the acquisition frame, and rescaling to take into account the difference due to use of a neutral density filter for the acquisition frame, which reduces the flux by a factor 50. The contrast curves of the two individual observing runs were also calculated in the same way, in order to allow the study of possible differences in outcome due to atmospheric conditions or related effects.

\section{Results and discussion}
\label{sec_results}

In this section, we present the results and various considerations that go into the determination of actual detection limits for $\epsilon$ Eri. First we show the actual final image, and corresponding contrast as a function of angular separation (Sect. \ref{sec_contrast}). We then continue our discussion from Sect. \ref{sec_obsstrat} about the grid calculations based on theoretical spectral models from the literature that were used to choose an optimal filter for the observations, and how (in combination with the measured contrast) they yield detection limits for the whole range of ages and masses allowed by the models (Sect. \ref{sec_comparison}). Since the age of the system is fundamental in constraining these detection limits, an attempt to constrain the age as far as possible based on the existing literature follows (Sect. \ref{sec_age}). The brightness contrast is subsequently translated into mass limits (Sect. \ref{sec_masslimits}). Further constraints are then added from independent dynamical data (Sect. \ref{sec_dyn}). Finally, all of these issues are combined to produce detection limits and constrain the allowed parameter range of any planets that may reside in the system (Sect. \ref{sec_limits}).

\subsection{Difference image and contrast}
\label{sec_contrast}

\begin{figure*}[htb]
   \centering
   \includegraphics[width=15cm]{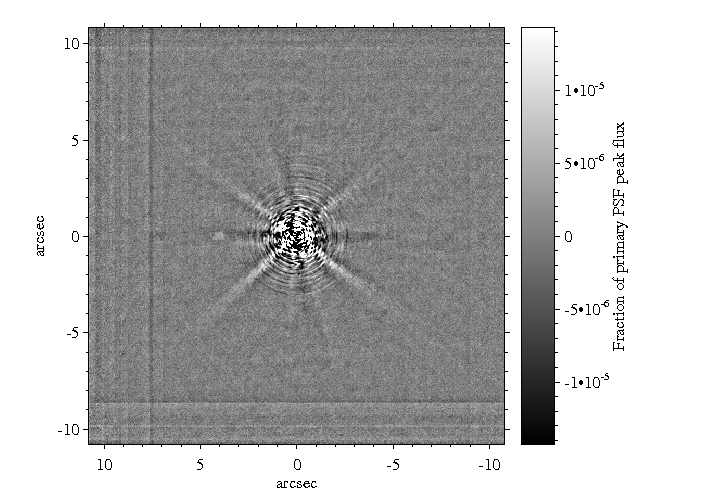}
\caption{Output difference image of $\epsilon$ Eri. The orthogonal line-shaped artifacts toward the edges of the image are due to the jittering, which shifts the field of view by a few arcseconds for each individual integration. In an unsaturated, undifferenced image, the PSF peak of $\epsilon$ Eri would here correspond to $3.5*10^5$ counts at a Strehl ratio of 85 \%. North is up and East is to the left in the image.}
\label{nbpic_large}
    \end{figure*}

\begin{figure}[htb]
   \centering
   \includegraphics[width=9.0cm]{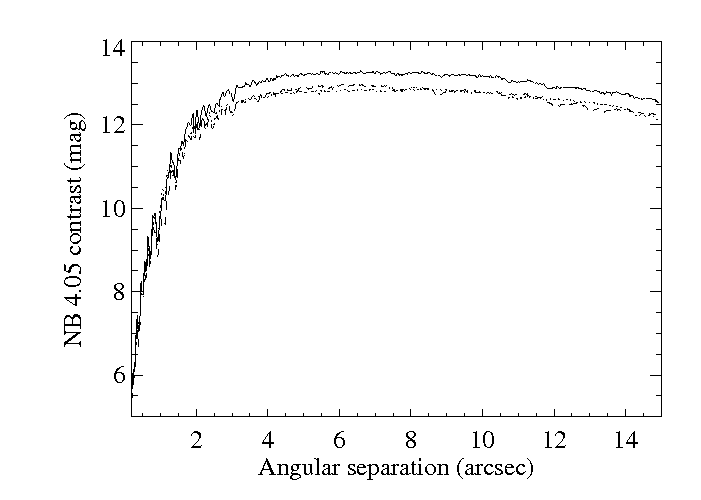}
\caption{Image contrast achieved in the NB4.05 filter with the ADI technique. Solid line: The full combined data set. Dashed and dotted lines: The two individual observing runs before co-addition. The decrease in contrast beyond about 10$''$ is due to the fact that the individual jittered frames overlap completely out to 10$''$, but non-completely outwards of this separation. Hence, the available effective exposure time decreases with separation in this range.}
\label{error_obslong}
    \end{figure}

\begin{figure}[htb]
   \centering
   \includegraphics[width=9.0cm]{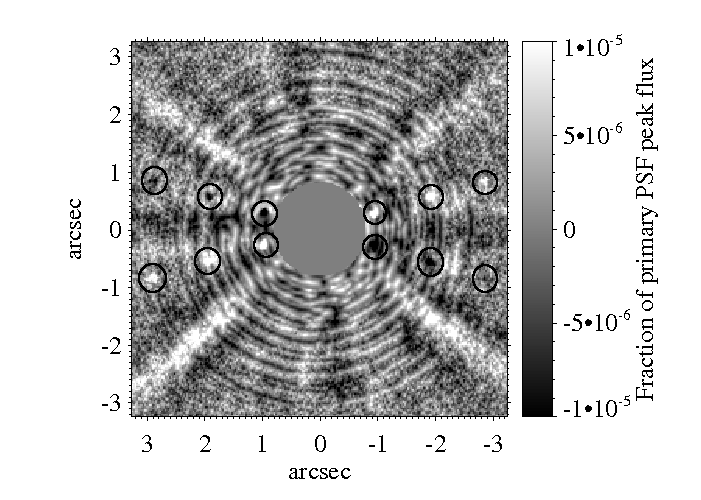}
\caption{Six artificial companions at 3$\sigma$ significance, two each at separations of 1$''$, 2$''$, and 3$''$. The center of the image has been artificially set to zero to enhance visibility. 3$\sigma$ is a useful constraint for most of the image range. However, in some cases it is problematic, such as for the rightmost dark signature, which by chance coincides with a local maximum in the residual stellar PSF and becomes difficult to distinguish. North is up and East is to the left in the image.}
\label{artificial}
    \end{figure}

\begin{figure}[htb]
   \centering
   \includegraphics[width=9.0cm]{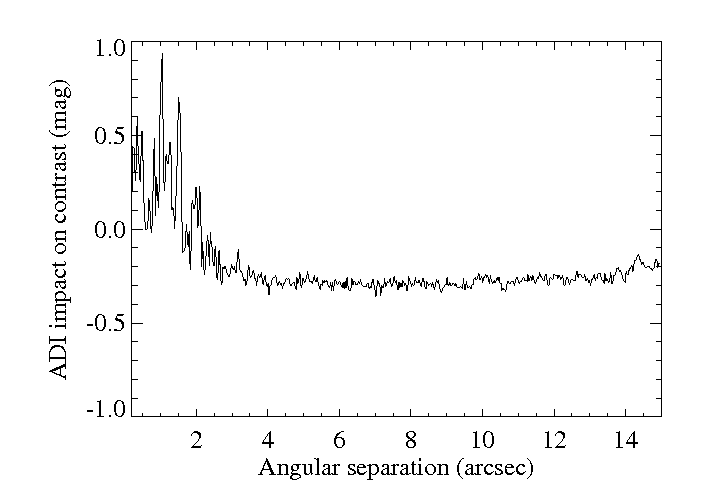}
\caption{The difference in performance between when ADI is used and when it is not. A positive number means that the contrast has increased when using ADI, for a single object signature. There is an apparent decrease in performance at large separations -- this is because there was no systematic noise in that range to be subtracted out, whereas random noise increased in the (negative) co-addition performed during ADI. However, at the same time, the number of object signatures is doubled, so there is in reality no decrease of performance in this range. The most substantial increase in performance is, as expected, at small angular separations, where systematic noise sources dominate.}
\label{adi_vs_not}
    \end{figure}

The final output image is shown in Fig. \ref{nbpic_large}. The remaining residuals in the image appear to be primarily due to rotation of PSF substructures between $0^{\circ}$ and $33^{\circ}$, and between the first and second observing run. Ideally, all the individual observations should be carried out at identical or symmetrical telescope orientations, for instance through symmetrical distribution of on-target time around the time of minimum zenith angle during a night. SDI data tends to efficiently mitigate effects such as spider rotation due precisely to the fact that the spectral image pairs are, by definition, taken at identical orientations. In any case, the achieved image contrast is excellent, as shown in Fig. \ref{error_obslong}. For instance, at an angular separation of 2.6$''$ we reach a contrast of $\Delta m_{4.05} = 13.2$ mag, which corresponds to a star-to-planet flux ratio of $1.6*10^5$. The contrast corresponds to the detectability of a companion by 3 $\sigma$. While this is not a very statistically stringent detection constraint for a single putative signature in the innermost part of the image where systematic noise dominates (see Fig. \ref{artificial}), it is stringent for the vast majority of the field where the residual noise is uncorrelated. The ADI technique in our application also yields two independent signatures, which increases the relevance of a putative detection. Thus we follow the previous convention for this technique and use 3 $\sigma$ as the criterion. The difference in performance between when ADI is used and when it is not (for a single object signature) is shown in Fig. \ref{adi_vs_not}.

Also shown in Fig. \ref{error_obslong} are the contrasts of the individual observing runs. The results from the individual runs are very similar, which is consistent with the similar Strehl ratios in the two cases. The contrast for the full dataset is significantly better than for the individual ones, implying that the residual noise is mostly incoherent, and that we benefit from using all the available data, which is not inherently obvious when the atmospheric conditions are so different. At $4.05$ $\mu$m, the isoplanatic angle is typically larger than $30''$, hence anisoplanatic effects are not expected to significantly affect the quality anywhere in the image range.

\subsection{Comparison with other techniques}
\label{sec_comparison} 

Detection limits for different methods in terms of planetary mass are derived in the following way: For each sampled angular separation and each filter, the measured contrast is compared with a physical contrast from the Baraffe et al. (2003) models, as described above. The comparison is made through linear interpolation between the grid points of the models to keep the result smooth -- for instance, if a 5 $M_{\rm jup}$ planet requires a contrast of 12 mag, and a 6 $M_{\rm jup}$ planet requires 11 mag, a measured instrumental contrast of 11.5 mag yields a detectable mass of 5.5 $M_{\rm jup}$. In the case of broad-band filters, this gives immediately the detectable mass at that separation. For narrow-band filters, we additionally need to apply the $\Delta _{\rm mag}$ correction using the Burrows et al. (2003) model. This has to be done iteratively, because a certain input mass gives a certain $\Delta _{\rm mag}$, which in turn gives a new mass, etc. After 4-5 iterations, convergence is reached and the mass limit thus acquired. 

For the case of SDI, we applied an additional correction. When qualitatively describing the process of SDI, it is often assumed that a planet with CH$_4$ absorption contains no flux within the $F3$ filter. In this way, when performing the $F1 - F3$ subtraction, it is easily understood that the stellar flux is canceled out, whereas the planetary flux is unaffected by the operation. This is, of course, a simplification, and not even a good approximation in many cases, such as for young and massive planets that do have significant flux in $F3$. One might think it more appropriate to regard the remnant flux as the difference between the individual $F1$ and $F3$ fluxes, but this is also an excessive simplification. What really happens during SDI is that the full $F1$ and $F3$ images are rescaled to a common $\lambda / D$-scale in order to optimize the subtraction of the primary PSF. As a part of this process, the $F3$ position of the prospective planet is shifted with respect to the $F3$ position by a factor $\lambda _3 / \lambda _1 = 1.625/1.575 \approx 1.03$. Hence, for a given angular separation $\rho$ between primary and secondary, the relative shift will be $1.03 \rho - \rho = 0.03 \rho$, so if $0.03 \rho \ll \lambda / D$, then most of the $F3$ flux from the hypothetical companion will be subtracted from its $F1$ flux. If, on the other hand $0.03 \rho \gg \lambda / D$, then $F3$ will yield an independent signature, such that is \textit{adds} to the achievable contrast. Around $\lambda = 1.6$ $\mu$m, the turnover point is on the order of $\rho _{\rm to} \approx 1.3''$ for the VLT. As implied, the physical reason for this is that beyond $\rho _{\rm to}$, there will be two independent signatures (one of positive counts, one of negative) of the companion in the final SDI image, separated by more than the FWHM of the PSF core. Hence for this special case, it is easier to detect those companions that do \textit{not} exhibit significant CH$_4$ absorption than those that do. In our simulations, we take this effect into account, considering both its dependence on angular separation and spectral energy distribution, with a correction factor applied to the contrast curve. The correction is calculated by generating two Gaussian (1D) functions with a FWHM set by $\lambda / D$, a relative shift given by $0.03 \rho$, and a relative amplitude fluxes in the $F1$ and $F3$ filters, respectively. The Gaussians are then subtracted, and the correction factor is calculated as the sum in quadrature of the negative and positive residuals over the absolute sum of the Gaussian corresponding to $F1$. $\Delta _{\rm mag}$ between $F1$ and $H$ is then calculated as described above.

Each of the calculations are made corresponding to a range of ages. The Burrows et al. (2003) models exist down to 1 $M_{\rm jup}$ for the ages of 100 Myr, 320 Myr and 1 Gyr. For these ages, the models are complete up to 10 $M_{\rm jup}$. Hence, we set the boundaries of our simulations to 0.1-1 Gyr and 1-10 $M_{\rm jup}$, since the Baraffe et al. (2003) models are also complete in this range. By complete, we mean that grid points exist at or around these boundaries such that no extrapolations need to be made.

\begin{figure}[htb]
   \centering
   \includegraphics[width=8.0cm]{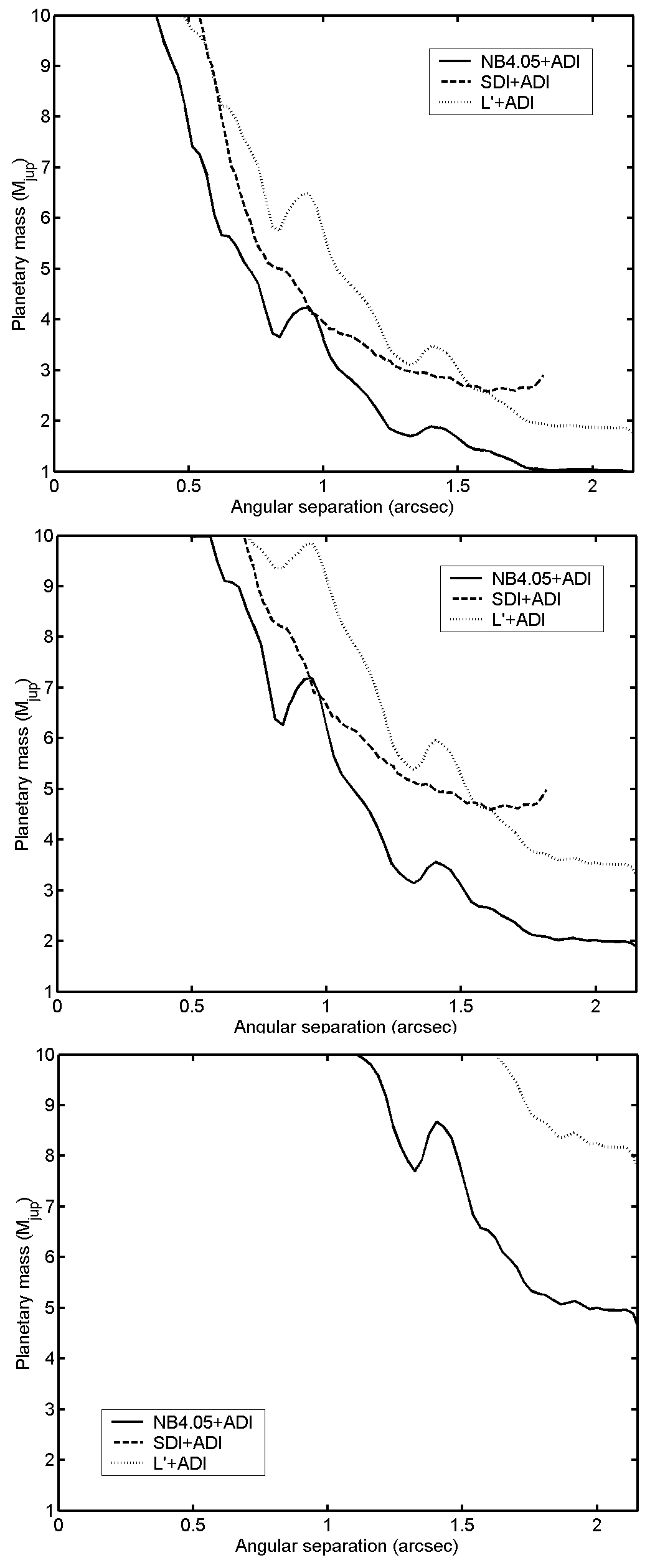}
\caption{Comparison of different high-contrast methods in terms of detectable mass of companions (in the simulated range of 1-10 $M_{\rm jup}$) to stars of ages 100 Myr (upper panel), 320 Myr (middle panel), and 1 Gyr (lower panel), and of similar brightness to $\epsilon$ Eri. Solid line: 4.05 $\mu$m with ADI. Dashed line: NACO-SDI with ADI. Dotted line: L'-band (assumed, see text) with ADI. Note that the NACO-SDI curve had about twice the integration time as the others. Under the assumption that $SNR \sim t^{1/2}$ (see Janson et al. 2007), this means that under the same integration time, the relative performance of NACO-SDI would get worse by 0.36 mag, generally corresponding to $\leq 0.5$ $M_{\rm jup}$ in these cases. Also note that L'-band is probably preferable to NB 4.05 in the background-limited regime, but the background limit is not yet reached at the outer cut-off of the figure (compare with Fig. \ref{error_obslong}).}
\label{cc_coll}
    \end{figure}

Performing all the calculations described above, for all relevant masses, separations, ages, methods, and wavelength ranges, we can calculate detection limits and compare different methods. While we have looked at several different filters available with NACO (e.g. broad-bands such as $H$ and $Ks$, and narrow-bands such as IB2.09 and NB2.17), here we focus on the three most interesting cases: NACO-SDI, L'-band and NB 4.05. In Fig. \ref{cc_coll}, we show a comparison between ADI in NB4.05 with NACO, and SDI+ADI around 1.6 $\mu$m with NACO-SDI, for the ages 100 Myr, 320 Myr, and 1 Gyr. Since the instrumental contrast curve for NB4.05 is jagged due to diffraction ring residuals, the mass limits that are based on this contrast become jagged as well, which makes the diagram more difficult to read out. For this reason, we have neighbour-averaged each separation point with the two nearest neighbours in each direction. This is a purely cosmetic operation which does not affect the factual content of the figures. Also shown is ADI corresponding to L'-band with NACO. In that case, the contrast is not based on a direct measurement, but on the assumption that the NACO ADI performance is equal in NB4.05 and L'-band (i.e., any difference is entirely due to the spectral energy distribution of a given object, which is generally favourable in NB4.05). This is a fair assumption, since the method and instrument is the same, though minor differences will exist since NB4.05 is not at the center of the L'-band, and thus the average Strehl ratio, background, and diffraction limit will be slightly different. The most important consequence of this result is that if the theoretical models are to be trusted, then ADI in NB4.05 significantly outperforms all other tested methods for all relevant masses and ages. This statement holds as long as the target has a similar near-infrared brightness as $\epsilon$ Eri. If the target is significantly fainter, the background will start to dominate, and eventually the contrast gained with respect to, e.g., NACO-SDI (which is not significantly affected by the thermal background at this level), will be void and the latter will be preferable. Another point that bears mentioning is that the results are specific to the AO system of NACO, and the Strehl ratio is one of the most important parameters governing the relative results at 1.60 and 4.05 $\mu$m. With a next-generation AO system, the Strehl ratio improvement at 1.60 $\mu$m can be expected to be large, whereas the Strehl ratio improvement at 4.05 $\mu$m, where the ratio is already very high, will be minor. For this reason, it is possible that SDI+ADI at 1.60 $\mu$m might still outperform ADI at 4.05 $\mu$m for the next-generation planet-finding instruments.

As a final point concerning the relative performance between NACO at 4.05 $\mu$m and NACO-SDI, we note that the field of view is considerable larger in the former case, and thus additionally better suited for the study of more distant companions. In the comparative plots, we have cut off the outer range in order to highlight the inner range, but note that NACO at 4.05 $\mu$m extends out to 10$''$ and beyond (see Fig. \ref{error_obslong}).

\subsection{Age of $\epsilon$ Eri}
\label{sec_age}

As we could see in Sect. \ref{sec_comparison}, the detectable mass is strongly dependent on the age of the system. Determining the age of young stars is however very difficult, and this is in particular true for the case of $\epsilon$ Eri. In fact, age estimations in the literature have been as disparate as the extent of our grid, about 0.1-1 Gyr. Here we will briefly examine this issue, to see if we can say something about the relative merits of different age estimations, and how far we can constrain the age range.

The methods that have been used to infer an age of $\epsilon$ Eri are large in amount: Isochronal dating, kinematics, Ca II  activity, X-ray activity, Li abundance, and rotation. Unfortunately, these methods give widely discrepant results, and many, if not all, have indications of being not well applicable for the particular case of $\epsilon$ Eri. In our discussion, we will focus on three papers that concern dating of $\epsilon$ Eri, Song et al. (2000), Fuhrmann (2004), and Barnes (2007), that cover all these methods. While several other papers exist that address the issue (e.g. Lachaume et al. 1999, Di Folco et al. 2004, Saffe et al. 2005), they cover the same methods and reach similar conclusions for each method, and the three former papers appear to be the most appropriate for age determination in this context.

Fuhrmann (2004) uses kinematics to associate $\epsilon$ Eri with the Ursa Major association, giving an age estimate of 200 Myr. Barnes (2007) uses gyrochronology (stellar rotation) to get an age estimate of 440 Myr. Song et al. (2000) ambitiously use five different methods (color-magnitude-diagram (CMD) with isochrones, Li abundance, X-ray emission, Ca II chromospheric activity, and young/old disk kinematics) and quote a final age range of 530 to 930 Myr. At a first glance, it may appear that the Song et al. (2000) results are far more secure, with five different methods consistently yielding an older age than the two other papers. However, we need to carefully check the applicability of each of the different methods. First of all, as is clearly noted also by Song et al. (2000), isochronal dating is not applicable to a target like $\epsilon$ Eri, because there is so little difference between different ages in this spectral type range beyond a few tens of Myrs. For instance, it is clear from the Song et al. (2000) CMD analysis that the age from this method is not only consistent with ZAMS, but it is also entirely consistent with the Pleiades empirical isochrone. Hence, the isochronal method does not constrain the age and is thus not useful for the analysis. The young/old disk kinematics analysis in the same paper identifies $\epsilon$ Eri as an old disk star, but it is very close to the border that separates the old and young disk, and the uncertainty of the identification procedure is not shown. Since the kinematic analysis of Fuhrmann (2004) identifies the spatial motion of $\epsilon$ Eri as being consistent with the Ursa Major association, and the analysis of that paper is more ambitious in this particular sense, we conclude that the kinematics of $\epsilon$ Eri does not exclude a younger age than the old disk population in general, and disregard this part of the Song et al. (2000) analysis. What remains are X-ray emission, Ca II activity and Li abundance, all of which are consistent with an age similar to the Hyades of 600-800 Myr. However, problems remain. Israelian et al. (2004) show that stars in the $\epsilon$ Eri spectral type range with detected exoplanets are significantly Li depleted compared to a reference sample. This result is confirmed by Gonzalez (2008), who finds the same trend also for rotation and activity. This implies that the values of Song et al. (2000), and even of Barnes (2007) should rather be seen as upper limits than realistic age estimations. Furthermore, there is a comment to be made with regards to the discrepancy between the age estimations from rotation (440 Myr) and activity (600-800 Myr). The physical link between activity as shown by Ca II and X-ray data is not well understood (Barnes 2007), but the activity is linked with the magnetic dynamo, which is in part governed by the rotation rate of the star. In this context, there is at least one obvious connection between age and activity: Stars are known to lose angular momentum with age, hence once they have reached the main sequence, the rotation rate decreases, the dynamo strength follows, and the activity decreases as a consequence. Note that in such a framework, activity is a less direct age indicator than rotation rate. This makes the former less reliable than the latter. This is more than a hypothetical line of reasoning, since it is known that activity cycles exist in solar-type stars on timescales over which the rotation rate can be approximated as being constant. Such cycles even include extended periods of inactivity -- for instance, during the period $\sim$1645-1715, the Sun showed almost no sunspots (the so-called 'Maunder minimum', after e.g. Maunder 1922). Meanwhile, if the rotation rate develops in a non-standard way, then within the same framework, both indicators would be unreliable.

Kinematic group membership as applied in Fuhrmann (2004) is a method that differs qualitatively from the other methods mentioned. If a given star is correctly identified as a member of a group, the age estimate is more precise than for any of the other methods, but on the other hand, if the star is misidentified and shares a common proper motion only by chance, it is irrelevant for dating purposes. The identification of $\epsilon$ Eri as an Ursa Major association member is not very convincing, as it is one of the most deviant stars in terms of spatial motion compared to the mean motion of those considered as group members. Thus, while we can't rule out $\epsilon$ Eri as an Ursa major association member, we also don't consider it a necessary match, and so we only use 200 Myr as a lower limit to the age range.


In summary, following the reasoning above, we consider the most probable age of $\epsilon$ Eri to be around 440 Myr (i.e. the age estimated from rotation rate in Barnes 2007), but we will use an age range of 200-800 Myr which corresponds to all the methods that we consider to be of some relevance for the age estimation.

\subsection{Mass limits}
\label{sec_masslimits}

\begin{figure}[htb]
   \centering
   \includegraphics[width=9.0cm]{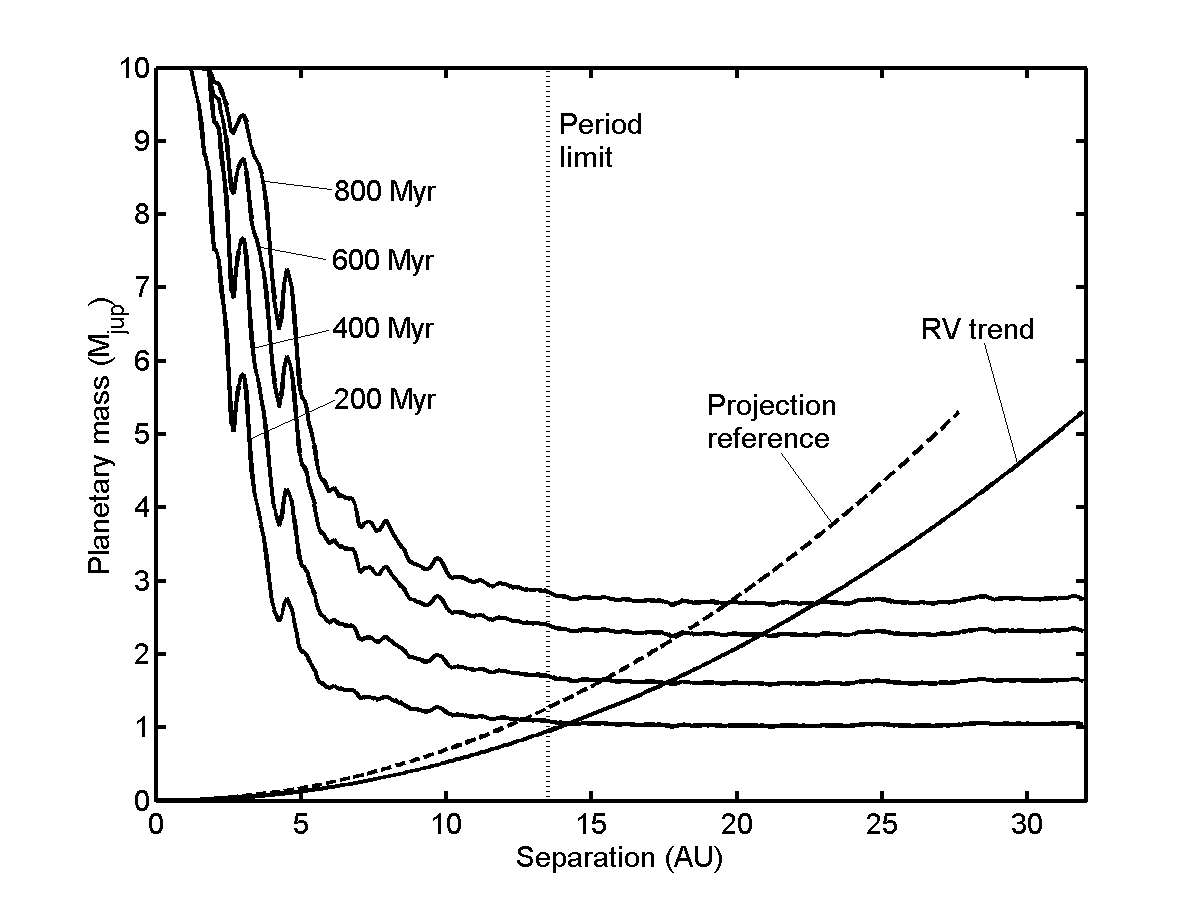}
\caption{Detection limits for ages of (from top to bottom) 800 Myr, 600 Myr, 400 Myr, and 200 Myr. Also plotted is the mass corresponding to the linear RV trend (solid line increasing with separation) given the $\epsilon$ Eri x hypothesis as described in Sect. \ref{sec_dyn}, and a vertical dotted line corresponding to the measured baseline of the trend. The dashed line is the minimum projected separation at $i = 30^{\circ}$. The latter curve serves to show that the detection limits are flat over the projection range for all ages, hence the projection is irrelevant and $\epsilon$ Eri x would simply be detectable at a given age from the separation where the solid RV line crosses the relevant detection limit, and outwards.}
\label{mlim_rv}
    \end{figure}

Using the simulations described in Sect. \ref{sec_comparison} and the age estimate from Sect. \ref{sec_age}, we get mass-equivalent detection limits as a function of physical separation shown in Fig. \ref{mlim_rv}. The radial velocity trend for which a corresponding curve is also shown in that figure is discussed in Sect. \ref{sec_dyn}, and the consequences for detection limits that arise from this will be discussed in Sect. \ref{sec_limits}.

\subsection{Input from dynamical methods}
\label{sec_dyn}

Direct imaging data can be combined with data from dynamical methods to constrain the presence of planetary mass companions. Radial velocity and astrometry data have already been used to infer the presence of $\epsilon$ Eri b (Hatzes et al. 2000 and Benedict et al. 2006). This input was used for the images in Janson et al. (2007) and is not a concern of this paper (the predicted projected separation is too small at the epoch of the NB4.05 images to improve on the SDI contrasts, see Sect. \ref{sec_limits}) -- here, we are instead primarily interested in what could be said about additional companions. In addition to these conventional methods, $\epsilon$ Eri offers an additional option to study dynamics in the system through its spatially resolved circumstellar debris disk, which can be expected to be shaped by interactions with any massive planets residing in the system. 

\subsubsection{Debris disk structure}
\label{sec_ddstructure}

Using numerical modeling, Quillen \& Thorndike (2002) find that the observed properties of the dust disk can be explained through the presence of a planetary body with approximate properties $M_{\rm pl} = 0.1$ $M_{\rm jup}$, $a_{\rm pl} = 41.6$ AU, and $e_{\rm pl} = 0.3$. In a more detailed set of simulations, Deller \& Maddison (2005) confirm that this is a feasible explanation, but claim that an additional, more massive planet is required further in to explain the inner cavity of the disk. This additional planet is not simulated in sufficient detail to give specific predictions, but the authors give a preliminary parameter range of 10-18 AU and a mass around 1 $M_{\rm jup}$. Hence, the result could be indicative of an additional planet candidate intermediate between the two other candidates ($\epsilon$ Eri b and $\epsilon$ Eri c). To simplify further discussion but simultaneously emphasize the speculative nature of this candidate, we will refer to it as $\epsilon$ Eri x. 

\subsubsection{Radial velocity}
\label{sec_radvel}

In the best RV fit of $\epsilon$ Eri b, there is a residual linear trend over the $\sim$27 year baseline, possibly indicative of a long-period physical companion. The trend is too large to be explained by secular acceleration (Benedict et al. 2006). It is also far too large to be caused by $\epsilon$ Eri c -- at a semi-major axis of $\sim 40$ AU, the orbital period would be $\sim 280$ years, and the resulting amplitude of the curve would correspond to a planetary mass of $\sim 8$ $M_{\rm jup}$, assuming that the orbital plane of the planet is the same as the plane of the disk, the rotational plane of the star, and the predicted orbital plane of $\epsilon$ Eri b, i.e. has $i \approx 30$ $^{\circ}$ (Saar \& Osten 1997, Greaves et al. 1998, Benedict et al. 2006). This is almost two orders of magnitude in mass above the Quillen \& Thorndike (2002) prediction, and should have been readily detected by previous imaging surveys (see Marengo et al. 2006). For $\epsilon$ Eri x, the inner possible range around 10 AU corresponds to a very short orbital period (35 years) with respect to the baseline of the linear trend (27 years), but the outer range of 15-18 AU yields reasonable periods of 65-85 years, and masses around 1-2 $M_{\rm jup}$, i.e. entirely consistent with the tentative values in Deller \& Maddison (2005). 

\subsubsection{Astrometry}
\label{sec_astrometry}

In Wielen et al. (1999), $\epsilon$ Eri was classified as a so-called $\Delta \mu$ binary candidate at the limit of detectability (roughly 3$\sigma$). $\Delta \mu$ binarity means that the short-term proper motion as measured by Hipparcos is different from the long-term proper motion given by the catalogues FK5 (the fifth Fundamental Catalogue) and GC (the General Catalogue), or a combination of the FK5 and GC positions with the Hipparcos position. The presence of a physical companion is a likely cause of such a trend, because an apparent short-term proper motion will include the astrometric effect of a sufficiently massive unseen companion, whereas over a long baseline of up to about 200 years such as in the FK5 case, a companion with an orbital period of a few decades or less will average out, and the measured long-term proper motion will correspond more strongly to the motion of the system as a whole. Since there is now a new data reduction made for Hipparcos data which significantly decreases astrometric residuals for bright stars like $\epsilon$ Eri (see e.g. van Leeuwen 2007), we have repeated the Wielen et al. (1999) analysis for this target with the new improved data.

We summarize the $\Delta \mu$ quantities along with the corresponding significance in Table \ref{nbtab3}. The reader is referred to Wielen et al. (1999) for specific definitions and discussion of the quantities involved -- we just briefly summarize the concepts relevant to our discussion in the caption of Table \ref{nbtab3}. For an intermediate orbit ($P \sim 30$ yr) as we are interested in here, $F_{\rm 0F}$ should be small, and the other values should be large in order to indicate (with a corresponding level of confidence) the presence of a companion. Note that $\mu _0$ and $\mu _{0(GC)}$ did not change when using the new reduction of the Hipparcos catalogue, since the positional difference between the old and new reduction is so small that it does not make a difference for the derived proper motion.

\begin{table}[htb]
   \centering
\caption[]{$\Delta \mu$ binary parameters with the old and new Hipparcos data reduction.}
         \label{nbtab3}
\begin{tabular}{lll}
            \noalign{\smallskip}
            \hline
            \hline
            \noalign{\smallskip}
Quantity                        & Old red.          & New red. \\
            \noalign{\smallskip}
            \hline
            \noalign{\smallskip}
$\Delta \mu _{{\rm FH},\alpha}$ [mas/yr]     & $1.70 \pm 1.08$  & $0.43 \pm 0.50$ \\
$\Delta \mu _{{\rm FH},\delta}$ [mas/yr]     & $3.31 \pm 1.05$  & $1.80 \pm 0.56$ \\
$F_{\rm FH}$                                 & $3.26$            & $3.40$           \\
            \noalign{\smallskip}
$\Delta \mu _{{\rm 0H},\alpha}$ [mas/yr]     & $1.76 \pm 1.02$  & $0.49 \pm 0.37$ \\
$\Delta \mu _{{\rm 0H},\delta}$ [mas/yr]     & $2.49 \pm 0.96$  & $0.98 \pm 0.35$ \\
$F_{\rm 0H}$                                 & $2.78$            & $3.28$           \\
            \noalign{\smallskip}
$\Delta \mu _{{\rm 0F},\alpha}$ [mas/yr]     & $0.06 \pm 0.54$  & $0.06 \pm 0.54$ \\
$\Delta \mu _{{\rm 0F},\delta}$ [mas/yr]     & $-0.82 \pm 0.60$  & $-0.82 \pm 0.60$ \\
$F_{\rm 0F}$                                 & $1.39$            & $1.39$           \\
            \noalign{\smallskip}
$\Delta \mu _{{\rm 0(GC)H},\alpha}$ [mas/yr] & $1.40 \pm 1.03$  & $0.13 \pm 0.39$ \\
$\Delta \mu _{{\rm 0(GC)H},\delta}$ [mas/yr] & $3.82 \pm 0.96$  & $2.31 \pm 0.38$ \\
$F_{\rm 0(GC)H}$                             & $3.96$            & $6.22$           \\
            \noalign{\smallskip}
            \hline
\end{tabular}
\begin{list}{}{}
\item[~] Note: $\Delta \mu$ is the difference between different proper motion values identified by the indices, where 'F' denotes FK5 proper motion, 'H' denotes Hipparcos proper motion, and '0' the change in position over baseline for FK5 and Hipparcos. 0(GC) is the same as the latter, but for GC instead of FK5. $F_{\rm FH}$, $F_{\rm 0H}$, $F_{\rm 0F}$ and $F_{\rm 0(GC)H}$ are the corresponding significance parameters, where $F_{\rm FH} = 3.44$ corresponds to a 99.7 \% significance.
\end{list}
\end{table}

It can be seen from Table \ref{nbtab3} that all the relevant significances, which were already high enough for Wielen et al. to consider $\epsilon$ Eri a limit-case $\Delta \mu$ binary, have increased with the new reduction. In addition, $F_{\rm 0(GC)H} = 6.22$ corresponds to a false alarm probability less than $4*10^{-9}$, indicating a strong detection of a companion. However, there are reasons to treat these numbers with great caution. In particular, note that in addition to decreased errors, the $\Delta \mu$ values have also decreased drastically -- this is why the FK5-based significance only increases marginally. It is also surprising that the $\Delta \mu _{\rm 0H}$ and the $\Delta \mu _{\rm 0(GC)H}$ are so different, since they should correspond to roughly the same observable, just for different long-term catalogs. For these reasons, we conclude that while there are indications from Hipparcos for a physical companion that could be compatible with the $\epsilon$ Eri x hypothesis, we emphasize that it is far from conclusive.

\subsection{Detection limits}
\label{sec_limits}

In Fig. \ref{mlim_rv}, we plot the expected mass of the hypothetical $\epsilon$ Eri x, under the assumption that it is responsible for the linear RV trend, and is on a circular orbit co-inclined with the rest of the system (including $\epsilon$ Eri b). Under these assumptions, we can exclude the $\epsilon$ Eri x hypothesis from 14 AU and outwards if the age is 200 Myr, and from 22.5 AU and outwards if the age is 800 Myr. Including also the conservative constraint that the period has to be substantially larger than twice the baseline, we find that if the system is 200 Myr old, $\epsilon$ Eri x cannot exist, whereas if it is older, there is an allowed parameter range between 14 and 22.5 AU, and between 1.0 and 2.7 $M_{\rm jup}$.

Unfortunately, exoplanet observations have shown that a perfectly circular orbit is not an ideal approximation -- many of the detected exoplanets have significantly eccentric orbits. Such orbits complicate the analysis due to their time-dependency in physical separation and orbital velocity. A planet on an elliptical orbit near periastron will reside further in towards the star for a given semi-major axis, and its mass corresponding to a given radial velocity amplitude will be smaller compared to a planet at the same physical separation and orbital phase, but with a circular orbit. This is due to the fact that the planet near periastron will move faster, such that a smaller mass is required to yield the same amplitude of the stellar reflex motion. However, a planet observed at a non-special (i.e., random) epoch will be unlikely to reside near periastron, because it spends a very small fraction of its orbit in that phase. In addition, the RV trend will change on a much faster timescale for a planet near periastron, such that to explain a linear trend with a baseline of 27 years, the full orbital period has to be several times larger, requiring a large orbit and thus also a large planetary mass. For a planet near apastron, the opposite is true. A planet on an elliptical orbit is likely to reside near apastron at a random epoch, and for a very eccentric orbit, the orbital period only needs to be a rather small amount ($<2$ times) larger than the observed baseline. On the other hand, the lower orbital velocity compared to a planet on a circular orbit with the same physical separation leads to a larger mass to yield the same velocity amplitude.

To estimate the impact of orbital eccentricity on our detection limits, we perform a simple simulation in the following way: We wish to simulate detection probability $\chi_{\rm det}$ as a function of semi-major axis $a$, and for this reason we step through 0.1 to 30 AU with steps of 0.1 AU. A representative eccentricity distribution is chosen by using the observed eccentricity distribution of RV-detected exoplanets (from the Extrasolar Planets Encyclopaedia, http://exoplanet.eu) outside of 0.1 AU where orbital circularization ceases to be important. This gives 179 realistic eccentricities. For each semi-major axis and eccentricity, we simulate 200 uniformly random orbits and instances, represented by the argument of periastron $\omega$ and the fraction of the period $f_{\rm P} = t/P$, where $0 < t < P$. Following the standard equations for Keplerian motion, this yields the eccentric anomaly $E$ (since the dependence $E(f_{\rm P})$ is non-linear, this is done numerically), which in turn yields the projected separation $\rho$. $\chi_{\rm det}$ is determined as

\begin{equation}
\chi _i =
\left\{\begin{array}{ccc}
1 & , & \textrm{~if~}M_{\rm rv}(a_i) / w_{i} \ge M_{\rm cc}(\rho _i) \\ 0 & , & \textrm{~otherwise~}
\end{array}\right.
\end{equation}

where $i$ is an index representing each simulated orbital instance, $M_{\rm cc}(\rho_{i})$ is the measured detection limit at the calculated projected separation and $M_{\rm rv}(a_{i})$ is the mass predicted by the RV trend for the $a$ at the given $i$ if the orbit was circular. $w_{i} = v_{i}/v_{0,i}$ is the fraction of real orbital velocity at the epoch corresponding to $i$ over the velocity of a circular orbit with semi-major axis $a_i$ -- i.e., it properly takes into account the eccentricity of the orbit. In addition to this, we must ensure that simulation $i$ is consistent with the baseline of observations. This is difficult to do stringently, as one would need to have specific information about the curvature of the (quasi-)linear trend, which would be to over-interpret the data at hand. In order to remain flexible with regards to curvature, we set the requirements that: 1) $P(a_{i}) \geq 2*27*w_{i}*q$ yr, and 2) $P(a_{i}) \geq 27*q$ yr, where $q$ is an arbitrary parameter governing which fraction of a period the baseline can correspond to. When $q=1$, the first condition claims that for a circular orbit, the baseline covers at least half the period (which is clearly as conservative as it gets), and that if the eccentricity is non-zero and the companion is close to periastron ($w_{i} > 1$), then the period must be larger because the observed trend represents a smaller fraction of the observed orbit, and the opposite holds true for apastron. The second condition at $q=1$ accounts for the fact that no matter how close to apastron a companion is, it clearly could never have a period shorter than the baseline itself (again, as conservative as it gets). By setting $q$ to some value $q \geq 1$, we can adjust our limits to be less conservative and, to some extent, more realistic. Here we will consider two different cases: $q=1$ and $q=2$.

\begin{figure}[htb]
   \centering
   \includegraphics[width=9.0cm]{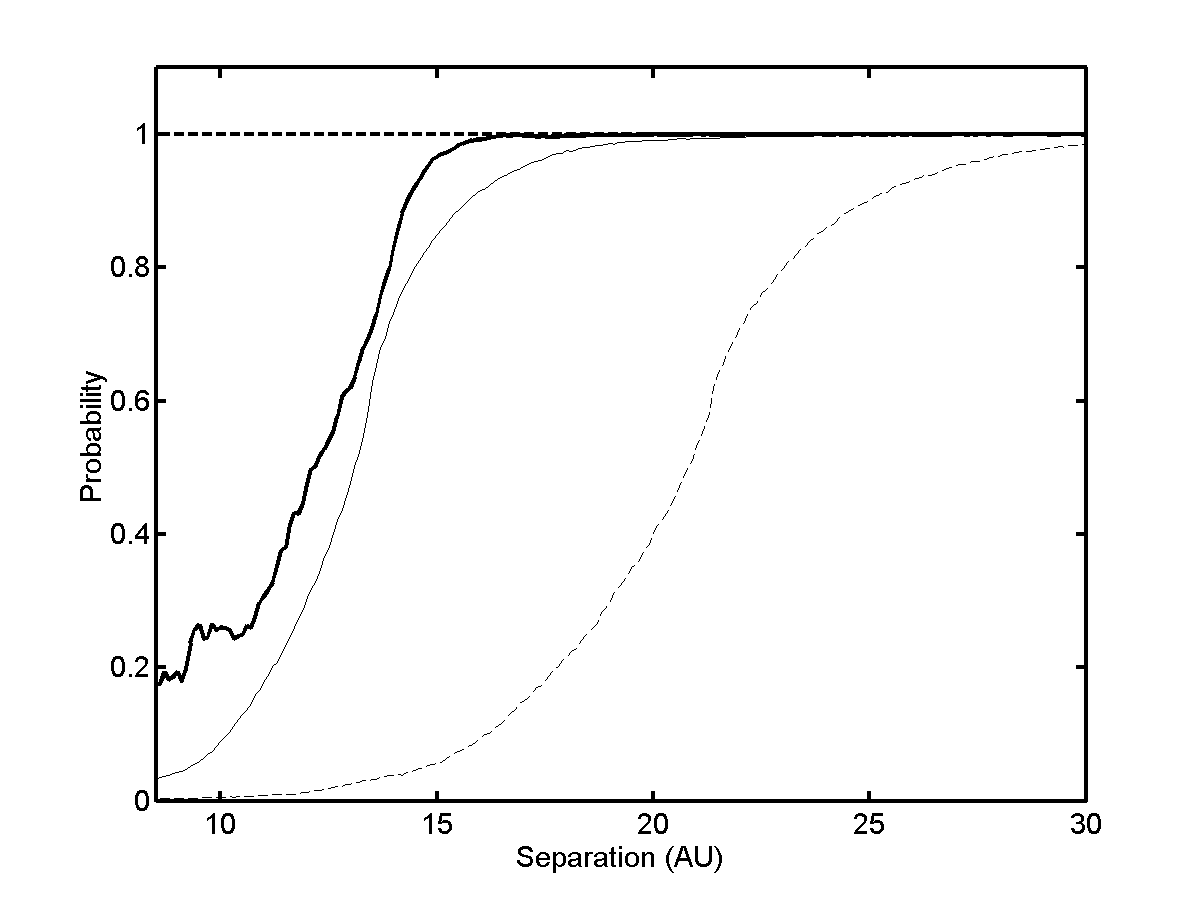}
\caption{Detection probability as a function of semi-major axis, at an age of 200 Myr. Thick lines: $\chi_{det}$ for $q=1$ (solid line) and $q=2$ (dashed line). Thin lines: Fraction of occurrences for a given semi-major axis that yield acceptable orbits, in the sense that the hypothetical planet could be consistent with the basic features of the RV trend (such as minimum baseline), for $q=1$ (solid line) and $q=2$ (dashed line). The meaning of the latter curves is that in the ranges where they give low probability values, it is unlikely that a planet could reside at those semi-major axes and still have an orbit that is consistent with the measured RV trend, according to the conditions specified by the $q$ values.}
\label{probfig_200}
    \end{figure}

\begin{figure}[htb]
   \centering
   \includegraphics[width=9.0cm]{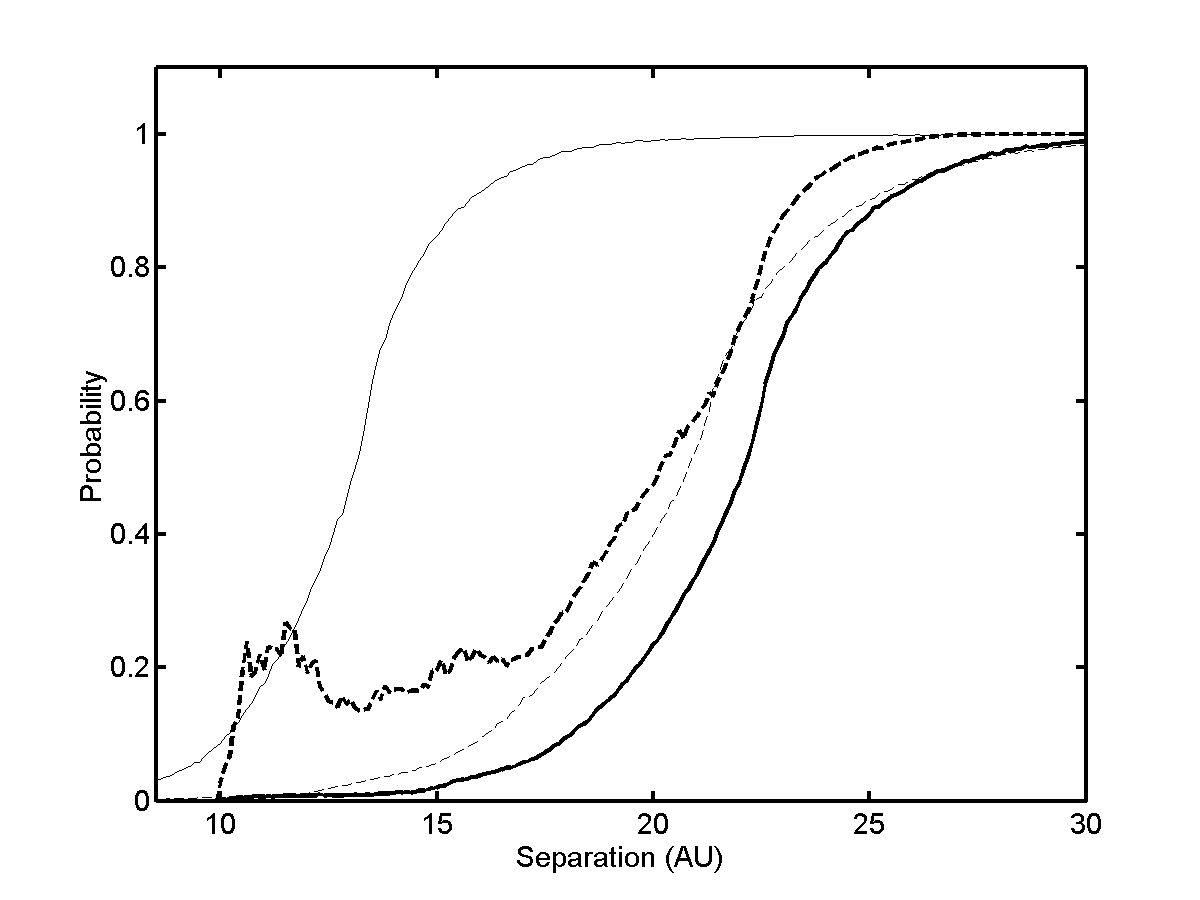}
\caption{Same as Fig. \ref{probfig_200}, but at an age of 800 Myr.}
\label{probfig_800}
    \end{figure}

Finally, we can calculate $\chi_{\rm det}$ as a function of $a_{i}$ as the mean of all $\chi_{i}$ corresponding to a given $a_{i}$. Only the $\chi_{i}$ that pass the period requirements are included in the mean. Separations $a_{i}$ that are small enough that no $\chi_{i}$ pass the criterion, such that no mean can be evaluated, are not included. We plot $\chi_{det}$ for the ages of 200 Myr and 800 Myr, and $q$-values of 1 and 2 in Figs. \ref{probfig_200} and \ref{probfig_800}. We see that there is now an allowed parameter range between 8.5 and 16.3 AU (for $\chi_{\rm det} < 0.95$) for $\epsilon$ Eri x even at 200 Myr for $q=1$, this is due to small orbits near apastron. We note that this parameter space is almost entirely due to the very conservative (or, from the perspective of allowed near-apastron orbits, very generous) period limits. If a more intuitive constraint such as $q=2$ is set, then $\chi_{det} = 1$ everywhere. For 800 Myr, the corresponding ranges are 8.5 to 27.3 AU for $q=1$ and 8.5 to 24.7 AU for $q=2$.

We note that regardless of whether there exists an $\epsilon$ Eri x or not, the RV trend serves as a useful indicator of what sort of companions could be detected in the RV data. For instance, returning to Fig. \ref{mlim_rv}, we see that at 8.5 AU where the 800 Myr contrast curve starts to be sensitive to 3 $M_{\rm jup}$ planets, the circular RV trend would correspond to less than 0.5 $M_{\rm jup}$, so if there was a 3 $M_{\rm jup}$ in the system at that separation, there would be an RV trend with a velocity amplitude 6 times larger than the existing linear trend, which would make the amplitude even larger than the one attributed to $\epsilon$ Eri b, and extend over a much longer baseline, making it very easy to discern. For smaller semi-major axes, the amplitude would be even larger. Thus since no such trend exists, we can combine the RV and imaging data to conclude that there are no companions to $\epsilon$ Eri more massive than 3 $M_{\rm jup}$. This is valid over all possible semi-major axes, since we are sensitive to 3 $M_{\rm jup}$ out to 14$''$, outward of which Marengo et al. (2006) provide a sensitivity to 1 $M_{\rm jup}$ companions. Fig. \ref{mlim_rv} corresponds to a circular orbit, but the conclusion can be extended to elliptical orbits, since the only way a planet can be more massive and still correspond to the same RV amplitude at a given semi-major axis is if it is near apastron, in which case it would be easier to detect in the imaging. 

Of course, there is a hypothetical possibility that a companion could exist that does not share an orbital plane with the rest of the system -- in that case, projection effects could still mask a massive companion in rare cases. However, for the case of disk-formed planets, which we would certainly expect to share an orbital plane with the circumstellar disk, we can conclude that whatever the actual number and orbital parameters of planetary companions to $\epsilon$ Eri are, they all must be less massive than 3 $M_{\rm jup}$.

\section{Conclusions}
\label{sec_conclusion}

We have performed a deep, high-contrast imaging observation of $\epsilon$ Eri with a new method that allows for detection of lower-mass planets at small separations from bright stars than previously possible. The method is based on ADI in the NB 4.05 $\mu$m filter, and compares favourably to other methods in the literature. Its advantage is however limited to very bright stars, since the background limit will dominate the contrast limit for faint stars, such that other methods (e.g. SDI+ADI or L'-band ADI) are preferable for such targets. As the final output images show no signs of any convincing companion candidate, we place constraints on any planets that may reside in the system. In particular, we test the hypothesis of the existence of an intermediate orbit planet possibly indicated by radial velocity, Hipparcos $\Delta \mu$ astrometry and numerical attempts to reproduce the circumstellar debris disk structure of $\epsilon$ Eri. In order to emphasize the hypothetical nature of such a companion, it is referred to as $\epsilon$ Eri x. We find that, due to the uncertain age of the $\epsilon$ Eri system, which we quote as 200-800 Myr based on a detailed investigation of the existing literature, the presence of $\epsilon$ Eri x can not be excluded based on our images, but its allowed parameter range can be constrained. In general, a combination of the existing RV data and high-contrast images allows us to conclude that whatever the amount and distribution of planets is in the $\epsilon$ Eri system, none of them can be more massive than 3 $M_{\rm jup}$.

\begin{acknowledgements}
The authors wish to thank Markus Kasper for useful discussion. M.J. gratefully receives financial support from IMPRS Heidelberg. The study made use of CDS and SAO/NASA ADS online services.
\end{acknowledgements}


\begin{thebibliography}{}

\bibitem[2008]{apai08} Apai, D., Janson, M., Moro-Martin, A. et al.\ 2008, ApJ 672, 1196

\bibitem[2006]{benedict06} Benedict, G.F., McArthur, B., Gatewood, G. et al.\ 2006, AJ 132, 2206

\bibitem[2003]{baraffe03} Baraffe, I., Chabrier, G., Barman, T.S., Allard, F., \& Hauschildt, P.H.\ 2003, A\&A 402, 701

\bibitem[2007]{barnes07} Barnes, S.A.\ 2007, ApJ 669, 1167

\bibitem[2007]{biller07} Biller, B., Close, L.M., Masciadri, A. et al.\ 2007, ApJS 173, 143

\bibitem[2003]{burrows03} Burrows, A., Sudarsky, \& D., Lunine, J.I.\ 2003, ApJ 596, 587

\bibitem[2005]{deller05} Deller, A.T. \& Maddison, S.T.\ 2005, ApJ 625, 398

\bibitem[2004]{difolco04} Di Folco, E., Thevenin, F., Kervella, P. et al.\ 2004, A\&A 426, 601


\bibitem[2004]{fuhrmann04} Fuhrmann, K.\ 2004, Astron. Nachr. 325, 3

\bibitem[2008]{gonzalez08} Gonzalez, G.\ 2008, MNRAS accepted

\bibitem[1998]{greaves98} Greaves, J.S., Holland, W.S., Moriarty-Schieven, G. et al.\ 1998, ApJ 506, 133

\bibitem[2000]{hatzes00} Hatzes, A.P., Cochran, W.D., McArthur, B. et al.\ 2000, AJ 120, 979


\bibitem[2006]{hinz06} Hinz, P.M., Heinze, A.N., Sivanandam, S. et al.\ 2006, ApJ 653, 1486


\bibitem[2004]{israelian04} Israelian, G., Santos, N.C., Mayor, M. \& Rebolo, R.\ 2004, A\&A 414, 601

\bibitem[2007]{janson07} Janson, M., Brandner, W., Henning, Th. et al.\ 2007, AJ 133, 2442

\bibitem[2008]{janson08} Janson, M., Brandner, W., \& Henning, Th.\ 2008, A\&A 478, 597

\bibitem[2007]{kasper07} Kasper, M., Apai, D., Janson, M., \& Brandner, W.\ 2007, A\&A 472, 321

\bibitem[1999]{lachaume99} Lachaume, R., Dominik, C., Lanz, T. \& Habing, H.J.\ 1999, A\&A 348, 897

\bibitem[2007]{lafreniere07} Lafreniere, D., Doyon, R., Marois, C. et al.\ 2007, ApJ 670, 1367

\bibitem[2003]{macintosh03} Macintosh, B., Becklin, E.E., Kaisler, D., Konopacky, Q., \& Zuckerman, B.\ 2003, ApJ 594, 538


\bibitem[2006]{marengo06} Marengo, M., Megeath, S.T., Fazio, G.G. et al.\ 2006, ApJ 647, 1437

\bibitem[1922]{maunder22} Maunder, E.W.\ 1922, MNRAS 82, 534

\bibitem[2005]{neuhauser05} Neuh\"auser, R., Guenther, E., Wuchterl, G. et al.\ 2005, A\&A 435, L13

\bibitem[2002]{quillen02} Quillen, A.C. \& Thorndike, S.\ 2002, ApJ 578, L149

\bibitem[2003]{saar03} Saar, S.H. \& Osten, R.A.\ 1997, MNRAS 284, 803

\bibitem[2005]{saffe05} Saffe, C., Gomez, M., \& Chavero, C.\ 2005, A\&A 443, 609


\bibitem[2000]{song00} Song, I., Caillault, J.-P., Barrado y Navascues, D. et al.\ 2000, ApJ 533, 41

\bibitem[2007]{vanleeuwen07} van Leeuwen, F.\ 2007, A\&A 474, 653

\bibitem[1999]{wielen99} Wielen, R., Dettbarn, C., Jahrei{\ss}, H., Lenhardt, H., \& Schwan, H.\ 1999, A\&A 346, 675

\end{thebibliography}
\end{document}